\documentclass[journal]{IEEEtran}
\usepackage{enumitem}   
\usepackage{mathtools}
\usepackage{amssymb}
\usepackage{amsmath} 
\usepackage{url}
\usepackage{graphicx}
\usepackage{stfloats}

\newtheorem{cor}{Corollary}[section]  
\newtheorem{thm}{Theorem}[section] 
\usepackage[]{algorithm2e}
\ifCLASSINFOpdf
\else
   \usepackage[dvips]{graphicx}
\fi
\usepackage{url}

\usepackage{graphicx}

\begin{document}

\title{A Fast and Simple Algorithm for computing the MLE of Amplitude Density Function Parameters}

\author{Mahdi Teimouri
\thanks{Mahdi Teimouri is with the Department of Statistics, Faculty of Science and Engineering, Gonbad Kavous University, Gonbad Kavous, 167, Iran (e-mail: teimouri@aut.ac.ir).}}
\markboth{Journal of \LaTeX\ Class Files, Vol. 14, No. 8, August 2015}
{Shell \MakeLowercase{\textit{et al.}}: Bare Demo of IEEEtran.cls for IEEE Journals}
\maketitle

\begin{abstract}
Over the last decades, the family of $\alpha$-stale distributions has proven to be useful for modelling in telecommunication systems. Particularly, in the case of radar applications, finding a fast and accurate estimation for the amplitude density function parameters appears to be very important. In this work, the maximum likelihood estimator (MLE) is proposed for parameters of the amplitude distribution. To do this, the amplitude data are \emph{projected} on the horizontal and vertical axes using two simple transformations. It is proved that the \emph{projected} data follow a zero-location symmetric $\alpha$-stale distribution for which the MLE can be computed quite fast. The average of computed MLEs based on two \emph{projections} is considered as estimator for parameters of the amplitude distribution. Performance of the proposed \emph{projection} method is demonstrated through simulation study and analysis of two sets of real radar data.    
\end{abstract}


\IEEEpeerreviewmaketitle

\section{Introduction}

\IEEEPARstart{I}{n} the context of radar clutter modelling, it is assumed that the total back-scattered complex signals recorded in a receiver follow a number of conditions. Suppose $X=X_{re} + j X_{im}$ denotes the total back-scattered complex signal that can be represented as 
\begin{align}\label{bscs}
X=\sum_{i=1}^{N} X_{i}\cos(\phi_{i}) + j \sum_{i=1}^{N} X_{i}\sin(\phi_{i}),
\end{align}
where $N$ is the number of received signals, $X_{i}$ is amplitude, and $\phi_{i}$ is phase of the $i$-th component. If the back-scattered complex signals follow some predetermined conditions, then a precise (efficient) statistical modelling of $X$ would be possible, see \cite{nikias1995signal,kuruoglu2004modeling,achim2006sar,karakucs2020modelling,ulaby2019handbook}. Among those conditions we mention: 
\begin{enumerate}[label=(\roman*)]
\item Total number of back-scattered complex signals must be large. 
\item Back-scattered complex signals must be independent. 
\item Back-scattered complex signals must be zero-mean with equal variances. 
\item Amplitude and phase of the back-scattered complex signals must be independent. 
\item The phase follows a uniform distribution on $(0,2\pi)$. 
\end{enumerate}

As pointed out by \cite{goodman1976some}, the real and imaginary parts of signal $X$ are independent and zero-mean so that, for a sufficiently large $N$, the central limit theorem guarantees that the real and imaginary parts of the total back-scattered complex signal $X$ follow asymptotically and independently a Gaussian distribution. Hence, distribution of the amplitude random variable $R=\bigl(X_{re}^{2} + X_{im}^{2}\bigr)^{1/2}$ becomes Rayleigh. Nevertheless, in some environments or situations, it is argued that some recorded signals do not follow some of the aforementioned conditions (i)-(v), see \cite{nikias1995signal}. For example, due to superposition of the impulsive echoes on each of terms in real (or imaginary) part of the total back-scattered complex signal $X$ represented in (\ref{bscs}), the real (or imaginary) part of $X$ no longer follows a Gaussian distribution, see \cite{nikias1995signal}. In such a case, although $X$ follows a zero-mean distribution, but likely heavy-tailed with infinite second moment. The statistical signal processing in presence of impulsive noises would be an important task since almost all of the parametric statistical approaches are susceptible to violation from underlying assumptions such as departure from Gaussianarity because of impulsive noises. Fortunately, in such a case, the generalized central limit theorem guarantees that the asymptotic (i.e. when $N$ in (\ref{bscs}) is sufficiently large) distribution of real and imaginary parts of the total back-scattered complex signal $X$ is $\alpha$-{stable}. In practice, it is recommended to assume that the real and imaginary parts of $X$ represented in (\ref{bscs}) follow independently a zero-location symmetric $\alpha$-{stable} (S$\alpha$S) distribution, see \cite{nikias1995signal,pierce1996rcs,tsakalides2001high}. It should be noted that there are a number of symmetric and heavy-tailed candidates for modelling the real and imaginary parts of the total back-scattered complex signal $X$ including, e.g. the Student's $t$ ($t$), generalized Gaussian (GG), etc. But, recalling from the generalized central limit theorem, the only choice for limiting distribution of the real and imaginary parts of $X$ is $\alpha$-{stable}.
\par 
There have been several efforts to model the radar clutter using $\alpha$-{stable} distribution. It sounds that \cite{tsihrintzis1997evaluation}, \cite{tsihrintzis1995signal}, and \cite{tsakalides1997new} are among the main researchers who make popular the $\alpha$-{stable} distribution for sea clutter modelling. In \cite{fiche2012rcs}, for modelling the scattering field of sea surface, it is shown that the $\alpha$-{stable} distribution is more appropriate model than the Weibull and ${\cal{K}}$-distributions. The $\alpha$-{stable} distribution outperforms the Rayleigh, log-normal, Rice and ${\cal{K}}$ distributions for target classification in radar, see \cite{li2006classifier}. Assuming that the image samples of a synthetic aperture radar (SAR) follow $\alpha$-{stable} distribution, the moment-based estimator of the amplitude probability density function (pdf) parameters is given by \cite{kuruoglu2004modeling}. The S$\alpha$S model is also applied as the background clutter in ultra-wide-band SAR, see \cite{banerjee2003statistical}. Based on several observations of clutter distribution, many applications use the $\alpha$-{stable} model in order to improve the performance of their desired application. For a survey on radar clutter modelling and target detection in SAR radars using $\alpha$-{stable} distribution we refer the reader, for example, to \cite{wang2008ship} and 
\cite{hao2023robust}, respectively.
\par
However, there have been some efforts that focus on other candidates rather than the $\alpha$-{stable} distribution. We refer the reader to \cite{how1997no}, \cite{baker1991k,eltoft1998non}, and \cite{angelliaume2019modeling} (for using the ${\cal{K}}$-distribution), \cite{ferrara2011generalized} (for using generalized ${\cal{K}}$ distribution), \cite{rosenberg2010analysis} (for using ${\cal{KK}}$ distribution), \cite{ishii2011effect,chen2013analysis}, and \cite{fernandez2015estimation} (for using Weibull distribution), \cite{chen2013analysis} and \cite{xing2009statistical} (for using log-normal  distribution), and \cite{angelliaume2019modeling} (for using Pareto distribution), among others. 

\section{The $\alpha$-stable distribution}
\subsection{Univariate $\alpha$-Stable Distribution}
The characteristic function (chf) of univariate $\alpha$-{stable} distribution is given by
\begin{eqnarray}\label{chf}
\displaystyle
\phi(t)=\left\{\begin{array}{c}
\displaystyle
\exp\Bigl\{-\lvert \sigma t \rvert^\alpha
\left[1-j\beta
\
\displaystyle
\mathrm{sign}(t)\tan\bigl(\frac {\pi \alpha}{2}\bigr)\right]\\+j t\mu \Bigr\},~~~~
\mathrm{{if}}
\
\alpha \ne 1,
\\
\displaystyle
\exp\Bigl\{-\lvert \sigma t \rvert \Bigl[1+j\beta
\
\displaystyle
\mathrm{sign}(t)\frac {2}{\pi}\log \lvert t \rvert \Bigr]\\+j t\mu \Bigr\},~~~~
\mathrm{if}
\
\alpha= 1,
\end{array} \right.
\end{eqnarray}
where $j^{2}=-1$ and $\text{sign(y)}$ is the well-known sign function that becomes -1, 0, and +1 for $y<0$, $y=0$, and $y\geq 0$, respectively. Herein, $\alpha \in (0,2]$ is called index of stability (or tail parameter) which determines the tail thickness of $\alpha$-stable pdf. In the Gaussian case ($\alpha=2$), the tail thickness of $\alpha$-stable pdf becomes the smallest and smaller values of $\alpha$ yield thicker tail for pdf of this distribution. Parameter $\beta \in [-1, 1]$ determines the degree of skewness. The values of -1, 0, and 1 for $\beta$ correspond to totally skeweness to the left, symmetry, and totally skeweness to the right for the pdf of this family, respectively. Parameter $\sigma \in \mathbb{R}^{+}$ plays the role of dispersion (scale) around $\mu$. For this family, $\mu \in \mathbb{R}$ is the location parameter. 
The mean does
not exist for $\alpha$-stable distribution when $\alpha \leq 1$ and if it does, then the mean of $\alpha$-stable distribution becomes $\mu$, but with infinite variance for non-Gaussian ($\alpha<2$) case. Except for three specific members, including L\'{e}vy ($\alpha=0.5$ and $\beta=1$), Cauchy ($\alpha=1$ and $\beta=0$) and Gaussian, the pdf of other members either do not have closed form or just can be expressed in terms of the special functions. Therefore, the class of $\alpha$-{stable} distributions is mathematically intractable and almost all of the corresponding analyses are carried out using the numerical tools. Let $S(\alpha, \beta, \sigma, \mu)$ denote the class of  
$\alpha$-stable distributions with chf (\ref{chf}). In the univariate case, there are two important subclasses of $\alpha$-stable family including the symmetric $\alpha$-stable (S$\alpha$S) and the positive $\alpha$-stable distributions. The chf of each zero-location S$\alpha$S (or $S(\alpha, 0, \sigma, 0)$) distribution is obtained by setting $\beta=0$ and $\mu=0$ in chf (\ref{chf}). The corresponding chf is given by
\begin{align} \label{chf1}
\phi(t)=\exp\bigl\{- \sigma^{\alpha} \vert t \vert^{\alpha}\bigr\}.
\end{align}
The known members of S$\alpha$S distribution with closed from density function are Cauchy and Gaussian. The support of the class $S(\alpha<1,1, \sigma, 0)$ is concentrated on the positive real axis, and so, this subclass sometimes is called the positive $\alpha$-stable (P$\alpha$S) distribution. An important member of the P$\alpha$S family is $S(\alpha/2,1, [\cos(\pi \alpha/4)]^{2/\alpha}, 0)$ whose cumulative distribution function (cdf) is given by
\begin{equation} \label{cdfp}
F_{P}(p \vert \alpha)=\frac{1}{\pi}\int_{0}^{\pi}\exp\Bigl\{-p^{-\frac{\alpha}{2-\alpha}}A(u)\Bigr\}du,
\end{equation} 
where $A(u)$ in the right-hand side of (\ref{cdfp}) is
\begin{equation} \label{A}
A(u)=\frac{\Bigl\{\sin\bigl[ \bigl(\frac{\alpha}{2}\bigr)u\bigr]\Bigr\}^{\frac{\alpha}{2}}\Bigl\{\sin \Bigl[\bigl (1-\frac{\alpha}{2}\bigr) u\Bigr]\Bigr\}^{1-\frac{\alpha}{2}}}{\sin (u)}.
\end{equation} 
\subsection{Family of Gaussian Scale Mixture Distributions} \label{gsm}
The class of Gaussian scale mixture (GSM) distributions is in fact a family of isotropic symmetric, elliptical, and independent multivariate distributions in which a positive random variable is multiplied with the covariance matrix of the multivariate Gaussian distribution. Let random vector ${\boldsymbol{Y}}=(Y_1,\cdots,Y_d)^{\top}$ follows a GSM distribution. We have
\begin{equation}\label{rep0}
{\boldsymbol{Y}}\mathop=\limits^d \sqrt{g(U)} \boldsymbol{Z}+\boldsymbol{\mu},
\end{equation}
where $\mathop=\limits^d$ denotes equality in distribution, $g(.)$ is a real positive function, $U$ is a positive random variable, and ${\boldsymbol{Z}}\sim {\cal{N}}_{d}({\bf{0}},\Sigma)$. Herein, we write ${\cal{N}}_{d}({\bf{0}},\Sigma)$ to show a $d$-dimensional Gaussian distribution with mean ${\bf{0}}$ and a ${d\times d}$ covariance matrix $\Sigma$.
\subsection{Family of Elliptical and Spherical $\alpha$-Stable Distributions} \label{subgaussian}
If in (\ref{rep0}) we let $U\sim$ P$\alpha$S and $g(u)=2u$, then $\boldsymbol{Y}$ follows an elliptical $\alpha$-stable (E$\alpha$S) distribution whose chf is \cite{nolan2013}:
\begin{equation}\label{ch2}
E\Bigl [\exp\bigl\{ j \langle {\boldsymbol{t}}, {\boldsymbol{Y}}  \rangle\bigr\}\Bigr]=\exp\Bigl\{- \bigl( {{\boldsymbol{t}}}^{\top} \Sigma {\boldsymbol{t}} \bigr)^{\frac{\alpha}{2}} + j \boldsymbol{\mu}^{\top}\boldsymbol{t} \Bigr\}.
\end{equation}
If in (\ref{rep0}) we let $\Sigma=\sigma \boldsymbol{I}_{d}$ in which $\boldsymbol{I}_{d}$ is a $d \times d$ identity matrix, then $\boldsymbol{Y}$ follows an isotropic $\alpha$-stable (I$\alpha$S) distribution with index of stability $\alpha$, scale parameter $\sigma>0$, and location vector $\boldsymbol{\mu} \in \mathbb{R}^{d}$. So, both of E$\alpha$S and I$\alpha$S families are in fact a GSM models. A special member of I$\alpha$S family is obtained when $\boldsymbol{\mu}=\boldsymbol{0}$. In this case the chf of I$\alpha$S family is given by
\begin{align} \label{chf4}
E\Bigl [\exp\bigl\{ j \langle {\boldsymbol{t}}, {\boldsymbol{Y}}  \rangle\bigr\}\Bigr] =\exp\Bigl\{- \sigma ^{\alpha}\bigl( \sum_{i=1}^{d} t^{2}_{i} \bigr)^{\alpha}  \Bigr\}.
\end{align}
A simple argument shows that the pdf of random vector $\boldsymbol{Y}$ following a $d$-dimensional zero-location I$\alpha$S distribution with chf (\ref{chf4}) is given by
\begin{equation}\label{pdf1}
f_{\boldsymbol{Y}}(\boldsymbol{y})=\frac{1}{(2\pi)^{\frac{d}{2}}\sigma^d}\int_{0}^{\infty}\frac{f_{U}(u)}{[g(u)]^{\frac{d}{4}}}\exp\Big\{-\frac{\sum_{i=1}^{d} y^{2}_{i}}{2\sigma^2g(u)}\Bigr\} du.
\end{equation}
In bivariate case, i.e. $d=2$, the pdf in (\ref{pdf1}) turns into 
\begin{equation}\label{pdf2}
f_{\boldsymbol{Y}}(\boldsymbol{y})=\frac{1}{2\pi \sigma^2}\int_{0}^{\infty}\frac{f_{U}(u)}{\sqrt{g(u)}}\exp\Big\{-\frac{y^{2}_{1} + y^{2}_{2}}{2\sigma^2g(u)}\Bigr\} du.
\end{equation} 
It has been argued in \cite{nikias1995signal} that the joint distribution of real and imaginary parts of the received total back-scattered complex signal is a bivariate I$\alpha$S with pdf given by (\ref{pdf2}). Hence, the chf for the joint distribution of $X_{re}$ and $X_{im}$ is given by \cite{kuruoglu2004modeling}:
\begin{align} \label{chf3}
E\Bigl [\exp\bigl\{j\bigl( t_1 X_{re}+t_2 X_{im}\bigr) \bigr\}\Bigr] =\exp\bigl\{- \sigma ^{\alpha}\bigl(t_{1}^{2} + t_{2}^{2} \bigr)^{\frac{\alpha}{2}}  \bigr\}.
\end{align}
Clearly, in the univariate case, the chf (\ref{chf3}) specializes to (\ref{chf1}).
\subsection{Amplitude distribution} \label{amplitude}
It has been shown in \cite{nolan2013} that pdf of the amplitude $R=\sqrt{X^{2}_{re} + X^{2}_{im}}$ random variable is 
\begin{align} \label{pdfr}
f_{R}(r\vert \Psi)=r \int_{0}^{\infty}\frac{f_{U}( u \vert  {\alpha})}{2\sigma^2 u} \exp\Bigl\{-\frac{r^2}{4\sigma^2 u}\Bigr\} du,
\end{align}
where $\Psi=(\alpha,\sigma)^{\top}$ is the amplitude distribution parameter vector and $f_{U}( u \vert \alpha)$ denotes the pdf of P$\alpha$S distribution whose cdf given by (\ref{cdfp}). Indeed, pdf of the amplitude distribution represented in (\ref{pdfr}) is a Rayleigh scale mixture model in which $U$ plays the role of mixing random variable. This means that
\begin{align}\label{rep1}
R\mathop=\limits^d 2\sigma \sqrt{U}W,
\end{align}
where $W$ follows a standard Rayleigh distribution with pdf
\begin{align*}
f_{W}(w)=2w \exp\bigl\{-w^2\bigr\}.
\end{align*}
Note that when $\alpha=2$, since each P$\alpha$S distribution degenerates at point one, then pdf of the amplitude distribution given in (\ref{pdfr}) specializes to pdf of an ordinary Rayleigh distribution. Hence, for different values of $\alpha$, representation (\ref{pdfr}) produces a wide range of distributions. For this reason, distribution with pdf given by (\ref{pdfr}) is called sometimes the generalized Rayleigh in literature, see \cite{kuruoglu2004modeling}.
\section{The MLE for amplitude distribution}
Let $\hat{\Psi}=(\hat{\alpha},\hat{\sigma})^{\top}$ denote the MLE of $\Psi$. We have
\begin{align}\label{mle}
\hat{\Psi}=\underset{\Psi}{\operatorname{argmax}} \sum_{i=1}^{n} \log f_{R}(r_i \vert \Psi),
\end{align}
where $f_{R}(. \vert \Psi)$ is given by (\ref{pdfr}). Indeed, finding $\hat{\Psi}$ through (\ref{mle}) is computationally cumbersome. Let ${\cal{U}}(0,2\pi)$ account for the family of uniform distributions on $(0,2\pi)$. Here, we propose a method to obtain $\hat{\Psi}$ that is very fast. To do this, Theorem \ref{thm1} plays the main role. 

\begin{thm}\label{thm1}
Let transformation $\boldsymbol{T}=(T_1,T_2)^{\top}$ denote the \emph{projected} vector of the amplitude random variable $R$ in which $T_1=R \cos \theta$ and $T_2=R \sin \theta$. We have
\begin{align}
\boldsymbol{T} \mathop=\limits^d \boldsymbol{X},
\end{align}
where $\theta \sim {\cal{U}}(0,2\pi)$ and $R=\sqrt{X^{2}_{1},X^{2}_{2}}$ following a distribution whose pdf is given by (\ref{pdf2}).
\end{thm}
{\bf{Proof:}} is simple and is omitted for the sake of saving space.
\begin{cor}\label{cor1}
Both marginals of \emph{projection} (transformation) $\boldsymbol{T}$, i.e. $R\cos \theta$ and $R \sin \theta$, follow a zero-location S$\alpha$S distribution admitting representation given by
\begin{align}\label{rep2}
T_1=R\cos \theta &\mathop=\limits^d \sigma \sqrt{g(U)} Z,\\
T_2=R\sin \theta &\mathop=\limits^d \sigma \sqrt{g(U)} Z,
\end{align}
where $g(u)=2u$, $U\sim$ P$\alpha$S $\sim S(\alpha/2,1, [\cos(\pi \alpha/4)]^{2/\alpha}, 0)$, and $Z\sim {\cal{N}}(0,1)$.
\end{cor}
Corollary \ref{cor1} states that observing amplitude data $\boldsymbol{r}=(r_1,\dots,r_n)^{\top}$, then parameters $\alpha$ and $\sigma$ can be estimated through the \emph{projections} of each SGS model that admits representation (\ref{rep0}). To do this, the observed vector $\boldsymbol{r}$ must be \emph{projected} on horizontal ($\boldsymbol{r}\cos \boldsymbol{\theta}$) and vertical ($\boldsymbol{r}\sin \boldsymbol{\theta}$) axes where elements of random vector $\boldsymbol{\theta}=(\theta_1,\dots,\theta_n)^{\top}$ follow independently ${\cal{U}}(0,2\pi)$. Suppose $\hat{\Psi}_c=(\hat{\alpha}_c, \hat{\sigma}_c)^{\top}$ and $\hat{\Psi}_s=(\hat{\alpha}_s, \hat{\sigma}_s)^{\top}$ denote the MLE of parameters $\alpha$ and $\sigma$ based on \emph{projections} $\boldsymbol{r}\cos \boldsymbol{\theta}$ and 
$\boldsymbol{r}\sin \boldsymbol{\theta}$, respectively. The estimator $\hat{\Psi}=(\hat{\alpha},\hat{\sigma})^{\top}$ can be obtained as $\hat{\Psi}=(\hat{\Psi}_c+\hat{\Psi}_s)/2$. The process of \emph{projection} can be repeated for $N\geq 1$ times in order to reduce the bias and standard error of the constructed $\hat{\Psi}$. Since $\hat{\Psi}$ is computed by \emph{projecting} the amplitude data on the horizontal and vertical axes, the proposed approach in this study can be seen as a \emph{projection} algorithm. This algorithm is given by the following.
\begin{algorithm}
\caption{Projection algorithm for computing $\hat{\Psi}$.}\label{alg1}
 \KwData{$\boldsymbol{r}=(r_1,\dots,r_{n})^{\top}$\;}
 \KwResult{ $\hat{\Psi}$\;}
{\bf{Initialization:}} set $j= 1$, ${\text{S}}_{\alpha}=0,{\text{S}}_{\sigma}=0$, and $N=20$\;
 	{
 		\While{$j \leq N$}
		{
    			 sample $\boldsymbol{\theta}=(\theta_1,\dots, \theta_n)^{\top}$ where $\theta_{i}$s independently follow ${\cal{U}}(0,2\pi)$\;
			compute $\boldsymbol{t}_1=\boldsymbol{r}\cos(\boldsymbol{\theta})$ and $\boldsymbol{t}_2=\boldsymbol{r}\sin(\boldsymbol{\theta})$\;
			since $\boldsymbol{t}_1=(t_{11},\dots, t_{1n})^{\top}$ follows a zero-location S$\alpha$S distribution, compute the MLE of $\hat{\alpha}_c$ and $\hat{\sigma}_c$\;
			since $\boldsymbol{t}_2=(t_{21},\dots, t_{2n})^{\top}$ follows a zero-location S$\alpha$S distribution, compute the MLE of $\hat{\alpha}_s$ and $\hat{\sigma}_s$\;
set ${\text{S}}_{\alpha}= {\text{S}}_{\alpha} + (\hat{\alpha}_{c}+\hat{\alpha}_{s})/2$\;
set ${\text{S}}_{\sigma}= {\text{S}}_{\sigma} + (\hat{\sigma}_{c}+\hat{\sigma}_{s})/2$\;
   			$ j = j +1$\;
  		 }
return $\hat{\Psi}=\bigl({\text{S}}_{\alpha}/N,{\text{S}}_{\sigma}/N\bigr)^{\top}$ as the MLE of ${\Psi}$.
}
\end{algorithm}
\section{Data Simulation and model validation}\label{sec4}
This section has two parts. First, in order to demonstrate the performance of the proposed \emph{projection} method, we perform a simulation study. Second, the proposed \emph{projection} method is tested by applying it to two sets of real radar data.
\subsection{Simulation Study}
\begin{figure*}
{\includegraphics[width=44mm, height=30mm]{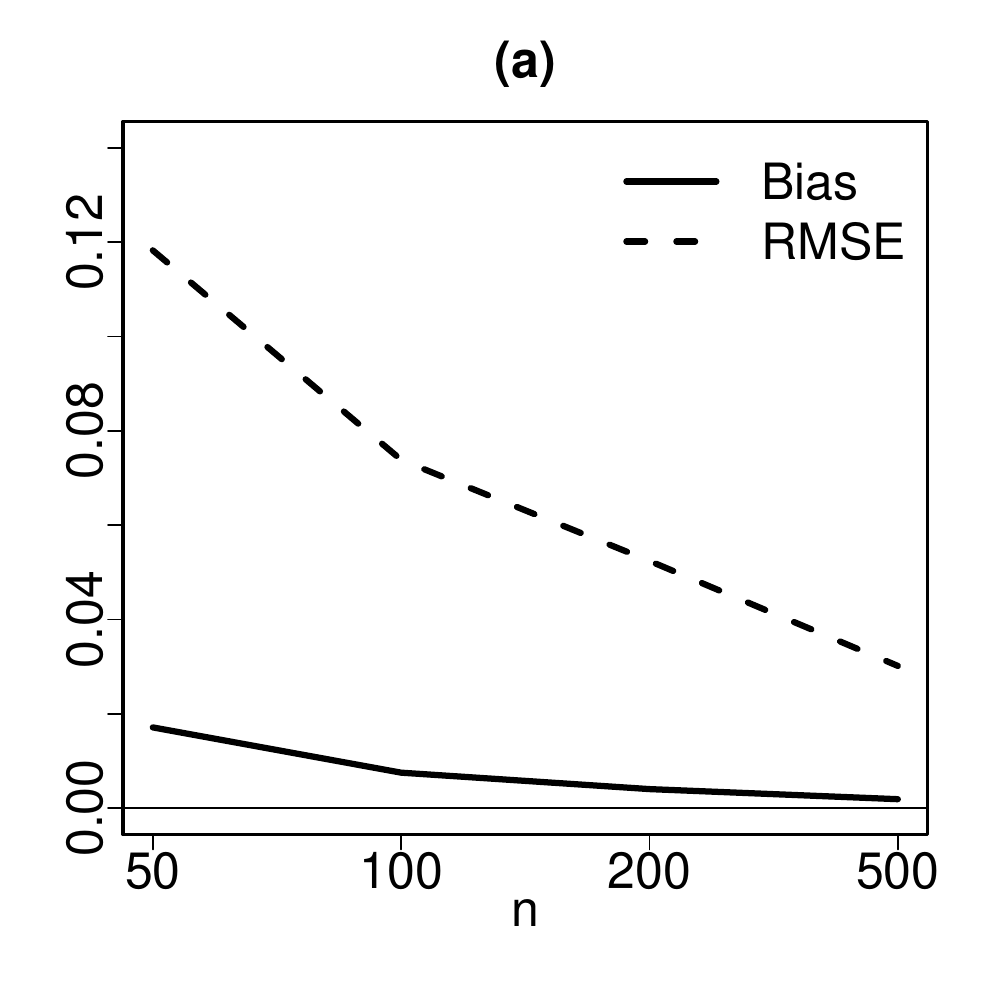}}
{\includegraphics[width=44mm, height=30mm]{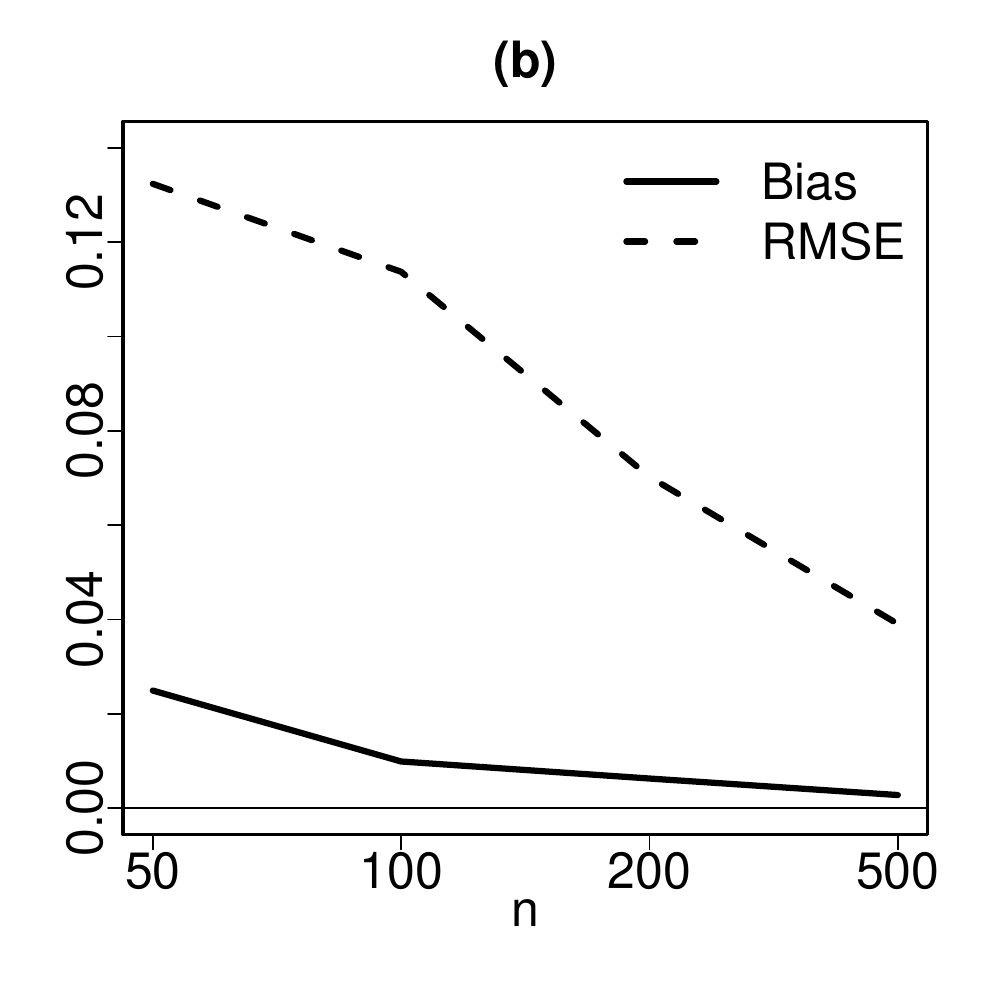}}
{\includegraphics[width=44mm, height=30mm]{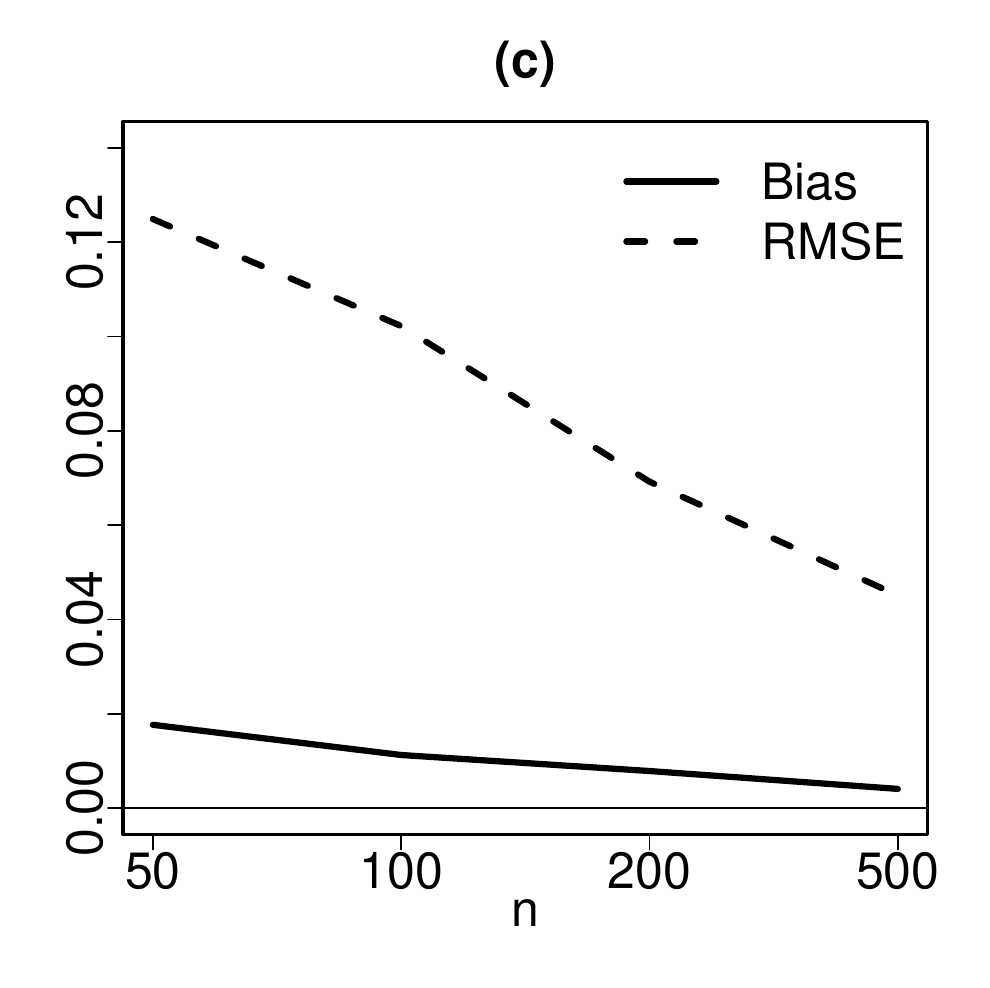}}
{\includegraphics[width=44mm, height=30mm]{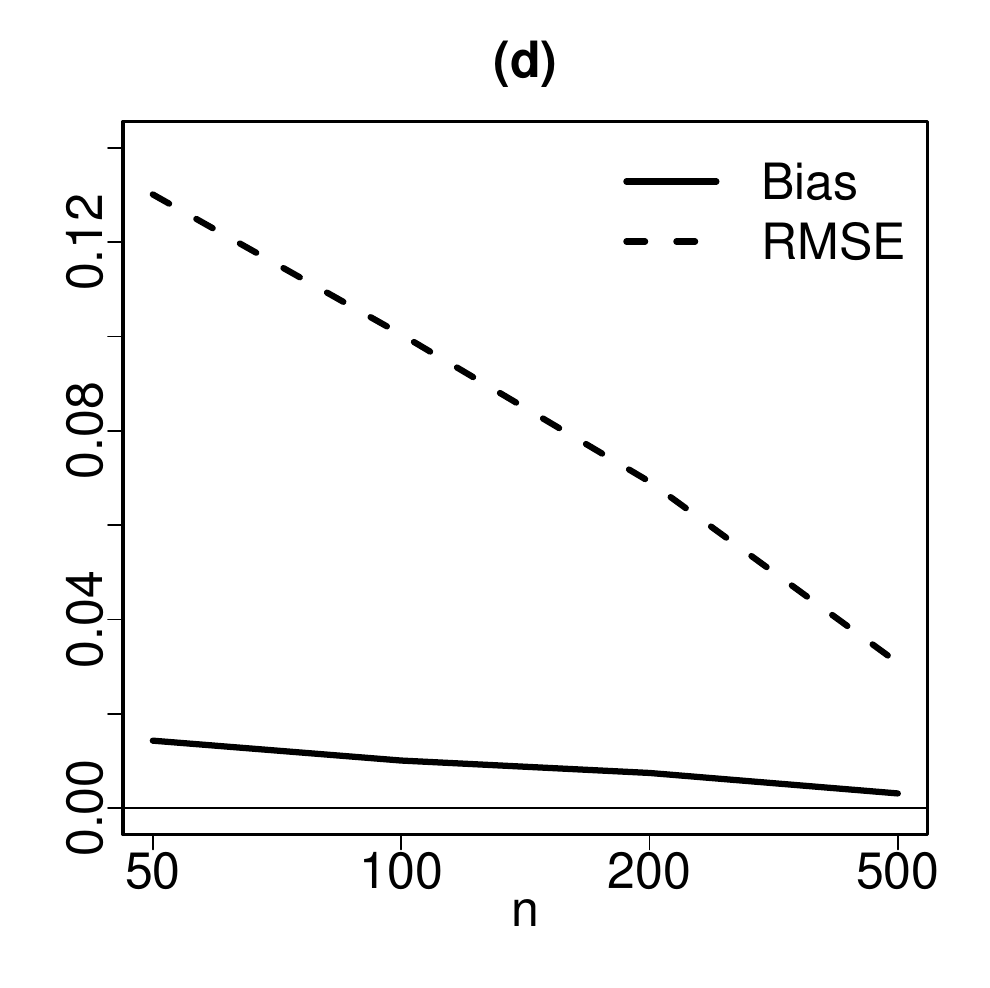}}
{\includegraphics[width=44mm, height=30mm]{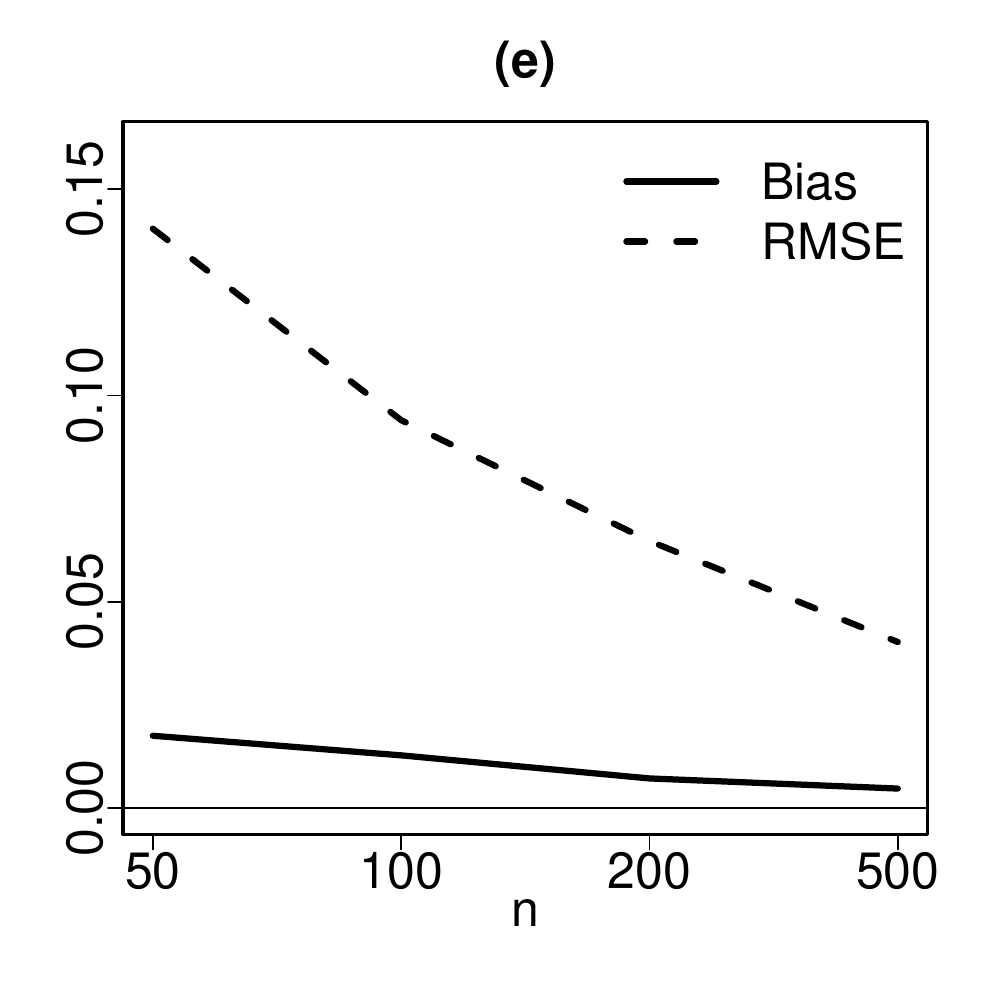}}
{\includegraphics[width=44mm, height=30mm]{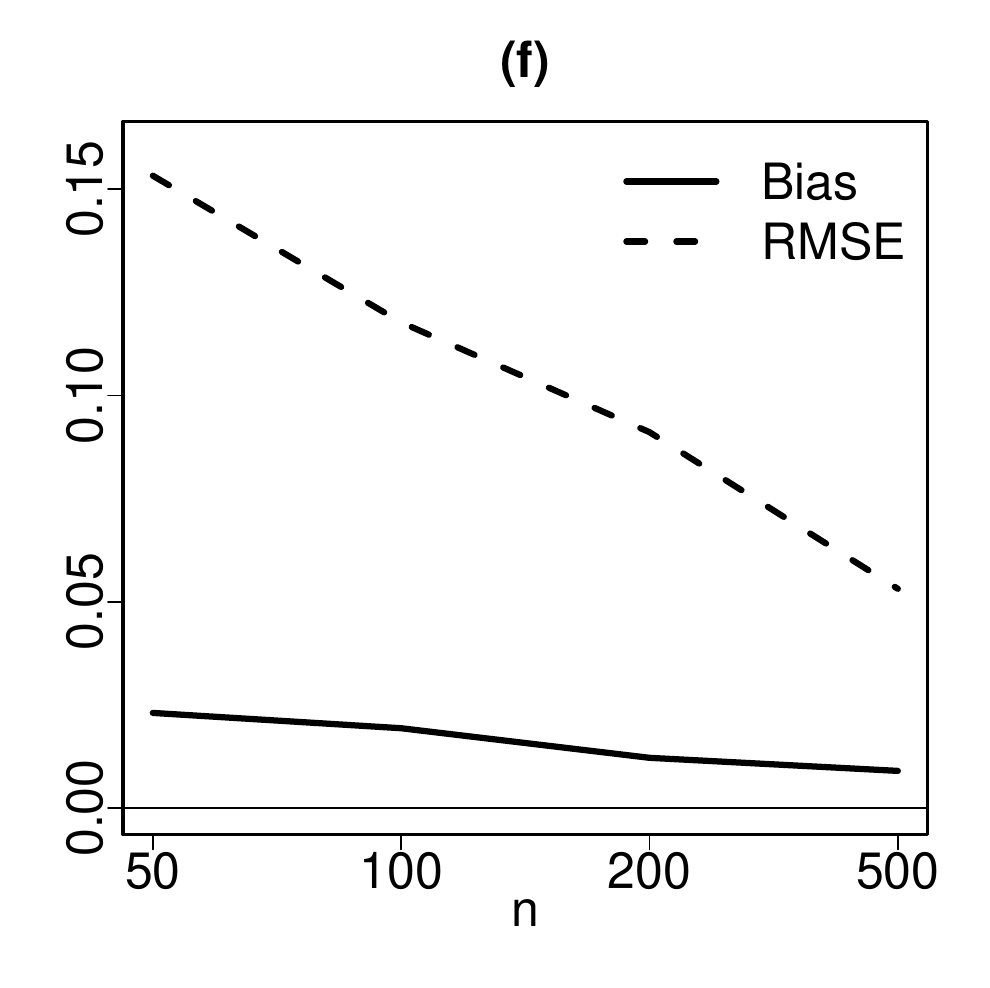}}
{\includegraphics[width=44mm, height=30mm]{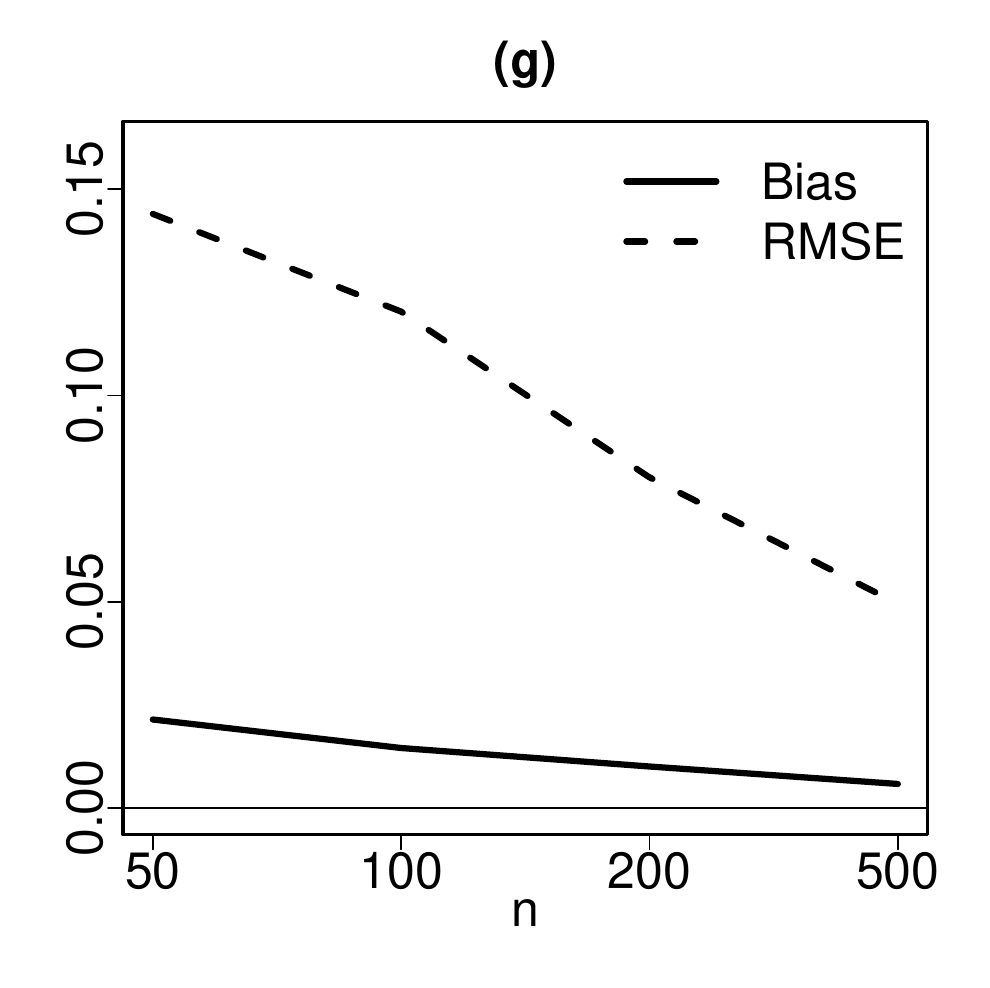}}
{\includegraphics[width=44mm, height=30mm]{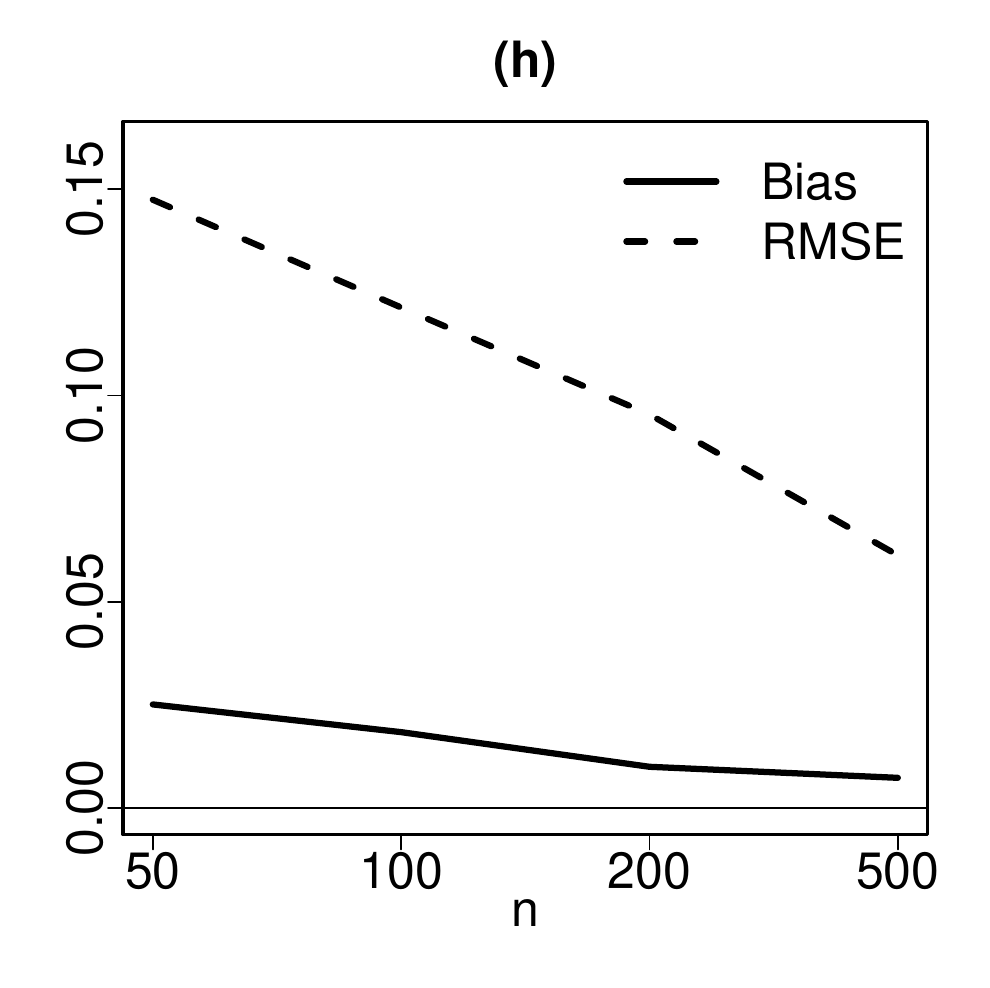}}
\caption{Bias and RMSE of $\hat{\alpha}$ for different values of $\Psi$. These include: (a) when $\Psi=(1,0.5)^{\top}$, (b) when $\Psi=(1.25,0.5)^{\top}$, (c) when $\Psi=(1.50,0.5)^{\top}$, (d) when $\Psi=(1.75,0.5)^{\top}$, (e) when $\Psi=(1,2)^{\top}$, (f) when $\Psi=(1.25,2)^{\top}$, (g) when $\Psi=(1.50,2)^{\top}$, and (h) when $\Psi=(1.75,2)^{\top}$.}
\label{fig1}
\end{figure*}

The bias and root mean squared error (RMSE) of the proposed MLE were computed. To this end, it was assumed that $\alpha$ takes values 1, 1.25, 1.5, and 1.75. Furthermore, $\sigma$ was fixed at 0.5 and 2.
For each setting of $\alpha$ and $\sigma$, the amplitude vector $\boldsymbol{r}$ was generated  for $M=1000$ trials of samples each of size $n=$50, 100, 200, and 500. The bias and RMSE criteria for $\hat{\alpha}$ were obtained as follows.
\begin{align*}
{\text{Bias}}&=M^{-1}\sum_{k=1}^{M}\bigl(\hat{\alpha}_{k}-\alpha\bigr),\\
{\text{RMSE}}&=(M-1)^{-1}\sum_{k=1}^{M}\bigl(\hat{\alpha}_{k}-\alpha\bigr)^2,
\end{align*}
Herein, $\hat{\alpha}_{k}$ denotes the MLE of $\alpha$ based on $k$-th sample of size $n$. Likewise, above criteria are defined for $\hat{\sigma}_{k}$. Fig. \ref{fig1} displays the Bias and RMSE of $\hat{\alpha}$ for different settings of the
$\alpha$, $\sigma$, and $n$. 
\subsection{Model Validation Using Real Data}
The proposed \emph{projection} method for computing the MLE of amplitude distribution parameters has been tested on a subset of real radar data \cite{farina1997high}. From the original dataset, 200 realizations have been selected. The results then were compared with the logarithmic moment estimator (LME) developed by \cite{kuruoglu2004modeling} and \cite{nolan2013}. Table \ref{tab1} shows the results of applying the LME and LME to two sets of real amplitude data. The goodness-of-fit statistics including the Kolmogorov-Smirnov (KS), Anderson-Darling (AD), and Cram\'{e}r Von Mies (CVM) have been used for selecting one of two estimators and assessing how well an I$\alpha$S model fits to amplitude data. As Table \ref{tab1} shows, the MLE clearly outperforms the LME.
\begin{table}[!h]
\caption{Summary results of the estimators LME and MLE.}
\centering{
\begin{tabular}{cccccccc} 
\cline{1-8}
      &               &\multicolumn{2}{c}{Estimation}&& \multicolumn{3}{c}{Measure}\\\cline{3-4} \cline{6-8}
Data&Estimator & $\hat{\alpha}$& $\hat{\sigma}$&& KS& AD &CVM\\ \cline{6-8}
\hline
first       &  LME  &1.207& 0.030&&0.135& 5.742 &1.101\\
            &  MLE   &1.587& 0.039&&0.070& 3.153&0.227\\
\hline\hline		
second  &  LME  &1.284& 0.034&&0.122& 5.693&1.031\\
            &  MLE  &1.624& 0.042&&0.079& 3.342&0.204\\
\hline\hline		
\end{tabular}}
\label{tab1}
\end{table}   
\section{Discussion}
We emphasize that the proposed \emph{projection} method in this work for computing the MLE of amplitude distribution parameters can be applied easily when the joint distribution of the real and imaginary parts of total back-scattered complex signal is I$\alpha$S distribution. 
Any violation from conditions (i)-(v), mentioned in Introduction, would affect the performance of the proposed MLE in this work. For example, if $\boldsymbol{\mu}\neq \boldsymbol{0}$ or when scale parameters of real and imaginary parts of the total back-scattered complex signal are not equal, the proposed \emph{projection} method for finding the MLE will not be efficient. In this work, we did not consider the LME during simulation study since the proposed \emph{projection} method shows superior performance than the LME. Furthermore, since each \emph{projection} follows a zero-mean S$\alpha$S distribution, our proposed method can be applied simply to estimate parameters of the amplitude distribution using the chf-based \cite{kogon1998characteristic} and quantile-based \cite{mcculloch1986simple} methods developed in the literature for $\alpha$-stable distribution with chf given by (\ref{chf}). It should be noted that developing the latter methods for S$\alpha$S distribution is easier than the general $\alpha$-stable distribution.
Our study whose results are not given here reveals that the quantile-based and chf-based methods outperform the LME. In this study, we used the MLE because of its asymptotic properties and efficiency. The proposed \emph{projection} method in this work for computing the MLE can be applied  very fast using available packages such as STABLE available at \url{http://www.robustanalysis.com}. The STABLE  has been developed for \verb+R+, \verb+Matlab+, \verb+Excel+, and \verb+Mathematica+. Moreover, the usage time for implementing the MLE based on $N=20$ \emph{projections} for real data in Table \ref{tab1} were about 1.14 seconds. The mixing variable $U$ in GSM model (\ref{rep0}) can be used for different distributions. For example, the Student's $t$ (for $g=1/u$), and generalized Gaussian (for $g=2^{\frac{1}{\alpha}}(2u)^{-1}$) \cite{teimouri2022maximum} distributions are among the well-known GSM moddels for which the mixing distributions are gamma and polynomially tilted positive $\alpha$-stable (PT$\alpha$S) \cite{devroye2009random}, respectively.
\par

\section{Conclusion}
We have proposed a method for computing the maximum likelihood estimator (MLE) of the amplitude distribution parameters whose real and imaginary marginals follow symmetric $\alpha$-stable (S$\alpha$S) distribution. The proposed method in this study is quite fast and works by \emph{projecting} the amplitude data into the horizontal and vertical axes through transformations $T_1=R\cos \theta$ and $T_2=R \sin \theta$, respectively. It has been proven that both \emph{projections} $T_1$ and $T_2$ follow S$\alpha$S distribution for which the MLE can be computed very fast. The performance of the proposed method has been demonstrated through simulation and modelling two sets of real radar data. The proposed \emph{projection} method in this study can be extended and then applied to find the MLE of the amplitude distribution parameters when joint distribution of real and imaginary parts of the total back-scattered complex signal admits a Gaussian scale-mixture model. 
\bibliographystyle{IEEEtran}
\bibliography{ref}
\end{document}